# Reactive Molecular Dynamics Simulation of a Buckybomb

Vitaly V. Chaban,[1] Eudes Eterno Fileti,[2] and Oleg V. Prezhdo[3]

[1] MEMPHYS - Center for Biomembrane Physics, Syddansk Universitet, Odense M., 5230, Kingdom of Denmark

[2] Instituto de Ciência e Tecnologia, Universidade Federal de São Paulo, 12231-280, São José dos Campos, SP, Brazil

[3] Department of Chemistry, University of Rochester, Rochester, NY 14627, United States

**Abstract**. Energetic materials, such as explosives, propellants and pyrotechnics, are widely used in civilian and military applications. Nanoscale explosives represent a special group because of high density of energetic covalent bonds. The reactive molecular dynamics study of nitro-fullerene decomposition reported here provides, for the first time, a detailed chemical mechanism of explosion of a nanoscale carbon material. Upon initial heating, $C_{60}(NO_2)_{12}$ disintegrates, increasing temperature and pressure by thousands of Kelvins and bars within tens of picoseconds. The explosion starts with $NO_2$ group isomerization into C-O-N-O, followed by emission of NO molecules and formation of CO groups on the buckyball surface. NO oxidizes into $NO_2$, and $C_{60}$ falls apart liberating $CO_2$. At highest temperatures, $CO_2$ gives rise to diatomic carbon. The study shows that the initiation temperature and released energy depend strongly on the chemical composition and density of the material. The established explosion mechanism provides guidelines for control of combustion and detonation on the nanoscale.

TOC Image

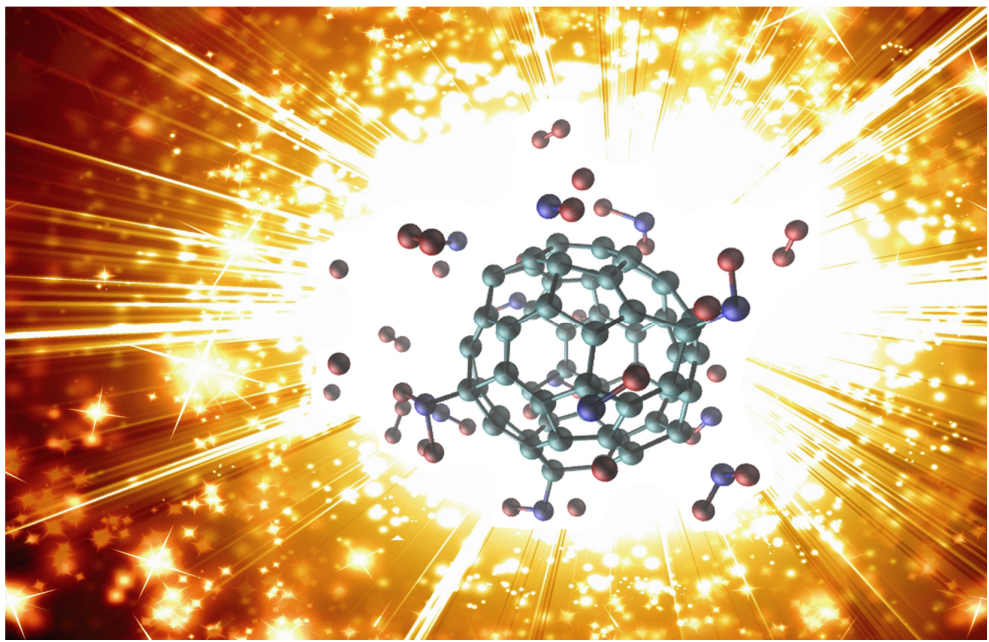

# 1. Introduction

Development of novel nanoscale materials has boosted a number of innovative research areas, including electronics, engineering, and biotechnology.[1-5] Carbon structures and high-energy systems constitute two particularly interesting classes of nanomaterials. The first class, which includes graphene, nanotubes and fullerenes, has been extensively studied during the last two decades. Applications of nanoscale carbon range from solar cells to anti-cancer therapy.[4-17] In turn, high-energy nanomaterials constitute a new promising field. Several important developments, such as nanoscale explosives, have been recently reported. The potential applications of nanoscale explosives extend beyond military purposes and industrial processes.[18-21]

Interplay of the unique properties of both classes – carbon structures and energetic nanomaterials – can lead to development of revolutionary and highly energetic carbon compounds. The field on nanoexplosives is fresh, requiring breakthrough developments of more efficient and less toxic materials. In this sense, carbonaceous materials exhibiting structural diversity, provide an exceptional platform for the development and research of reactive energetic systems.[2] Chemical methods have proven to be extremely efficient in changing the properties of fullerenes and nanotubes via covalent or non-covalent functionalization.[22,23] In particular, the horizon of the $C_{60}$ fullerene applications has been significantly extended by chemical modifications of its structure.[24]

Polynitrofullerenes are stable molecules with a high energy of formation. Their synthesis has been recently reported,[25,26] while prospective applications have been barely researched yet. The $NO_2$ group is well-known as a major source of oxygen, which contributes strongly to detonation and combustion processes via partial or full oxidation of intermediate products.[27] One would expect that nitrated fullerenes could emerge as promising nanoexplosives, whose power and sensitivity can be tuned.

## 2. Results and Discussions

The present work simulates a real-time explosion of dodecanitrofullerene, $C_{60}(NO_2)_{12}$, which we nickname a *buckybomb*. Reactive molecular dynamics is used to establish, for the first time, a detailed mechanism of explosion of a nanoscale carbon material. The study shows that upon initial heating to 1000 K, the $NO_2$ groups isomerize into C-O-N-O groups within a picosecond. Rapidly after that, NO molecules are emitted, and CO groups are formed on the buckyball surface. Finally, both NO and $C_{60}$ are oxidized. NO gas becomes $NO_2$ gas on a sub-10 ps time, and $C_{60}$ disintegrates producing $CO_2$. The $C_{60}$ disintegration continues for tens of picoseconds. The temperature and pressure grow by thousands of Kelvins and bars. At highest temperatures, 100 ps after the start of the explosion, $CO_2$ gives rise to diatomic carbon. The established time-resolved chemical transformations provide a unique perspective on the nanoscale explosion process, generating insights that can be used to design novel energetic materials.

The first nine $NO_2$ groups were attached to $C_{60}$ in such a way that every hexagon and pentagon contained a single $NO_2$ group. The additional three $NO_2$ groups were attached to random carbon atoms. The total number of $NO_2$ groups was selected to attain a balance between stability at room conditions and vigorous explosion upon initiation. Smaller numbers of $NO_2$ groups, in particular, six and nine, were preliminarily tested.

The list of the simulated systems is given in Table 1. Each system was simulated during 500 ps with an integration time-step of 0.1 fs. The explosion simulations were carried out in the constant energy ensemble (NVE), while the induced pressure was determined in the constant volume constant temperature (NVT) ensemble. Initial molecular configurations were generated using the PackMol[28] procedures to obtain system energies close to local minimum. NVE simulations were started at 1000 K, and system temperature was monitored until the explosive decomposed into individual atoms

(ca. 5000 K). 15 oxygen molecules were added to each system (Figure 1) to represent atmosphere, because it is anticipated to play an important role in the explosion kinetics.

Table 1. Simulated systems and representative results.

| Number of $C_{60}(NO_2)_{12}$ | Number of covalent bonds | Density, kg m$^{-3}$ | Time before decomposition, ps | Number of independent simulations |
|---|---|---|---|---|
| 1 | 386 | 190 | 70 | 3 |
| 2 | 769 | 320 | 90 | 7 |
| 4 | 1521 | 590 | 160 | 3 |
| 8 | 3028 | 1130 | 250 | 3 |

The algorithm of fragment recognition uses connection table and bond orders calculated at every time-step. The bond order cut-off used to identify molecular species is set to 0.3 for all bond types. Two fragments are considered separate molecules if all bond components, defined between them, exhibit orders smaller than 0.3. Note, that definition of a chemical bond is not unique, in principle. The selected value of the bond order cut-off influences the composition and concentration of intermediate products, but it does not influence the final (stable) products. Consequently, ReaxFF sporadically suggests existence of certain exotic molecules and fragments, which are not detected by any experimental technique, because of their transient nature and low stability. It is important to distinguish between bonded and non-bonded atom pairs in order to obtain translational kinetic energy, which is converted into temperature. The selected value (0.3) was tested in previous works, showing reliable and chemically sound results.

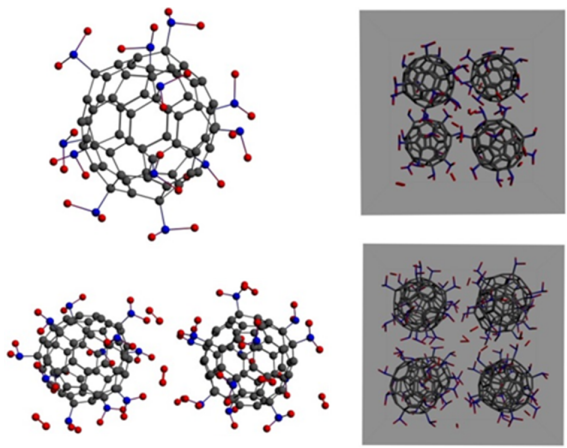

**Figure 1**. Starting configurations of the simulated systems. The unit box for the higher-density systems is depicted in gray. Carbon, nitrogen and oxygen atoms are gray, blue and red, respectively. Each system contains 15 $O_2$ molecules to represent an atmosphere. The $O_2$ molecules are made invisible in the isolated $C_{60}[NO_2]_{12}$ system for clearer representation of the explosive structure.

The non-equilibrium explosion dynamics were simulated in the following way. The potential energy of each system was minimized using the conjugate gradient algorithm for geometry optimization. The resulting configurations correspond to local energy minima in the absence of thermal motion. Afterwards, velocities corresponding to 1000 K, with respect to the initial number of covalent bonds (Table 1), were assigned. The classical equations of motion were propagated conserving the total energy. Figure 2 depicts evolution of temperature and pressure during the simulated explosions. Temperature evolution is shown up to 4000 K, since energy conservation at higher temperatures is unstable even with a 0.1 fs time-step. In turn, higher temperatures cannot be easily achieved upon real conditions, because of inevitable energy dissipation.

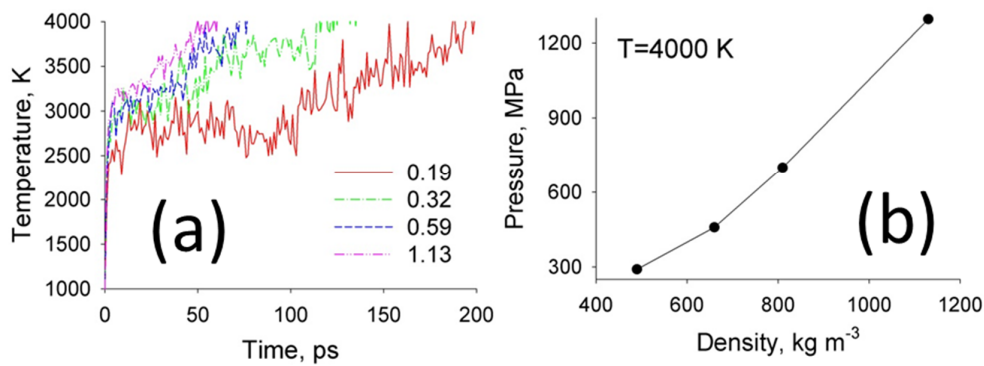

**Figure 2**. (a) Evolution of temperature upon explosion for various densities (see legend, in g cm$^{-3}$) of the $C_{60}[NO_2]_{12}$ explosive. (b) Pressure upon explosion for various densities of the $C_{60}[NO_2]_{12}$ explosive. Pressure is calculated at 4000 K, since explosion-like decomposition and drastic temperature increase are observed around this temperature.

For explosion to happen, an initiation event must take place. This event can include an exothermic reaction with a low energy barrier or mechanical impact. In case of a buckybomb, such reaction is isomerization of C-NO$_2$ groups into C-O-N-O groups (Figure 3). The reaction takes place within 1 ps and results in a temperature increase from 1000 to 2500 K (Figure 2). This stage does not depend on explosive density, since it is an intramolecular chemical reaction.

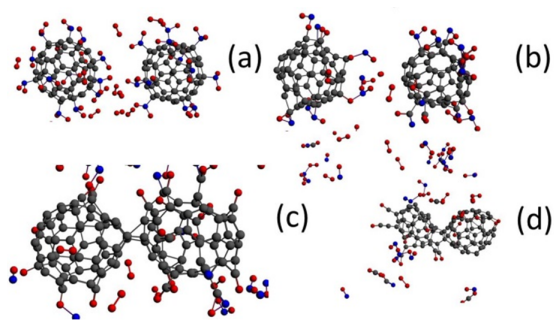

**Figure 3**. Mechanism of $C_{60}[NO_2]_{12}$ decomposition: (a) starting configuration; (b) isomerization of C-NO$_2$ groups into C-O-N-O groups; (c) formation of carbonyl groups at the buckyball surface, accompanied by NO molecule emission; (d) liberation of carbon

dioxide, transformation of nitrogen monoxide into nitrogen dioxide. Carbon, nitrogen and oxygen atoms are gray, blue and red, respectively. The box edges are omitted for clarity.

The temperature of 2500 K is sufficiently high to initiate partial decomposition of $C_{60}[NO_2]_{12}$. Certain fraction of NO and $NO_2$ molecules (Figure 4) leave the major particle, fostering formation of carbonyl groups, C=O. Additional C=O groups are formed at the buckyball surface due to interaction with molecular oxygen. The NO gas is unstable at 2500 K. It reacts with oxygen within very few time-steps to form $NO_2$, in agreement with experimental knowledge. Formation of carbonyl groups is the first step towards oxidation. This stage is fast (Figure 2), but less exothermic than the initiation stage involving $C-NO_2$ isomerization.

The carbonyl groups make buckyball highly unstable at ca. 3000 K. Significant fraction of carbon atoms break carbon-carbon bonds and join $CO_2$ molecules (Figure 4). A higher content of oxygen containing reactants in the system would increase the reaction rate even further.

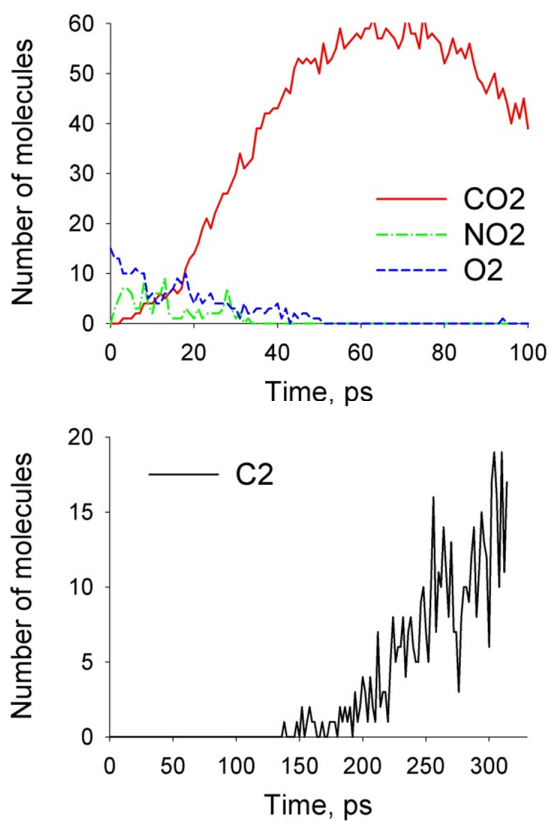

**Figure 4**. Evolution of the quantities of selected molecules ($CO_2$, $NO_2$, $O_2$, $C_2$) in the simulated systems. Most carbon atoms, released from partially or completely destructed buckyballs, create carbon dioxide gas. As explosion proceeds and system temperature increases, $CO_2$ molecules decompose giving rise to diatomic carbon.

Starting from ca. 4000 K, carbon-carbon bonds are broken quickly. This process releases the main portion of energy, which is responsible for a huge pressure elevation (Figure 2). The simulations were stopped at this stage, since drastic kinetic energy increase during the reaction makes the integration time-step of 0.1 fs insufficiently small to conserve the total energy. Therefore, temperature and pressure computed above the carbon-carbon bond decomposition temperature are insufficiently reliable. The system state at these very high temperatures is not reported.

The final mixture (Figure 5) is composed primarily of $CO_2$, $NO_2$, $N_2$ gases and linear carbon chains. NO and $O_2$ gases are absent, since NO was oxidized to $NO_2$ and $O_2$ formed carbon dioxide molecules with carbon atoms. The molecular nitrogen is at a lower state of internal energy than the oxides of nitrogen. This is why a significant number of nitrogen atoms exist as $N_2$ at the end of reaction. Interestingly, despite an excess of carbon atoms, $CO_2$ dominates over CO. Note, that the final mixture composition is dependent on the available atoms. For instance, the presence of hydrogen – whether in the form of molecular hydrogen or water vapor in the initial mixture of reactants – would result in formation of hydrocarbon chains. A larger fraction of oxygen molecules would result in carbon chain oxidation into carbon dioxide. This combustion reaction is known to be also exothermic.

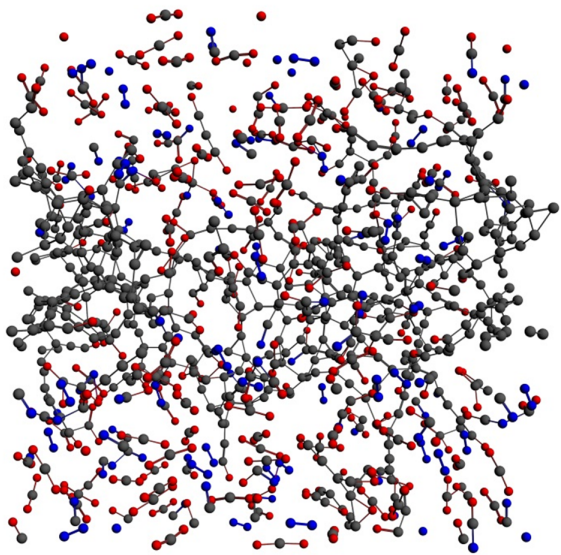

**Figure 5**. Molecular configuration around the time of explosion. The gases in this mixture generate pressure (Figure 3) that causes the explosive effect.

The time-step used for integration of equations of motion is a central parameter in constant energy MD simulations. It influences the validity of all results, being responsible for total energy conservation. The reactive MD method, in the form applied in this work, does not employ schemes for interaction energy cut-off, as is done customarily in conventional MD simulations. Therefore, energy conservation depends on the integration of Newton equations and accuracy of numerical procedures for energy calculation. Figure 6 provides a comparison for different time-steps. Simulation of reactions at the elevated temperatures requires significantly smaller time-steps than those usually applied in the non-reactive simulations (1-5 fs). Based on the performed analysis, we selected a 0.1 fs time-step, which achieves reasonable balance between accuracy and computation costs.

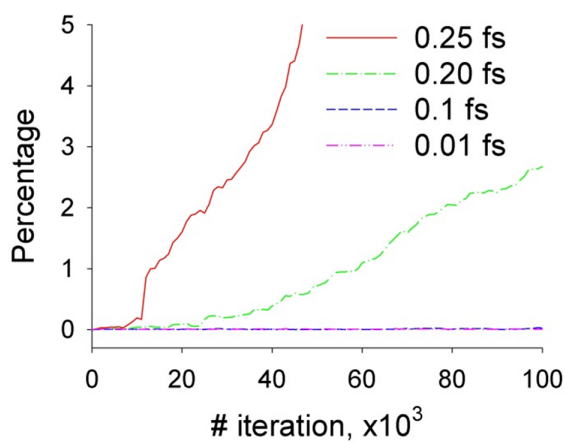

**Figure 6**. Violation of the total energy conservation in the $2 \times C_{60}(NO_2)_{12}$ system as a function of the propagation time-step. The calculation was started with the kinetic energy corresponding to 1000 K and minimized potential energy (optimized geometry).

3. Conclusions

To recapitulate, we have reported non-equilibrium reactive MD simulations of a $C_{60}[NO_2]_{12}$ buckybomb explosion. We have showed that this compound exhibits properties of a highly energetic material. Heated to 1000 K, $C_{60}[NO_2]_{12}$ decomposes spontaneously providing a significant amount of heat. The heat is released due to a high density of covalent energy stored by carbon-carbon bonds. $NO_2$ groups act as explosion initiators, providing the first portion of kinetic energy (temperature increase) to the system. Atmospheric oxygen actively participates in this reaction after the initial stage. The decomposition mechanism has been elucidated in detail. The rates of all elementary steps and the nature of the transient species have been established, the reaction products have been discussed. Fast liberation of chemical energy, resulting in a 3000 K temperature increase within 50-200 ps, provides exciting opportunities for chemistry and engineering.

**4. Methodology**

The molecular dynamics (MD) simulations were performed using quantum chemistry (QC) based reactive force field (ReaxFF). [29-32] This methodology was applied previously with success to address a number of complicated problems. [6,33-38] ReaxFF provides a nearly ab initio level of description of reactive potential surfaces for many-particle systems. The method treats all atoms in the system as separate interaction centers. The instantaneous point charge on each atom is determined by the electrostatic field due to all surrounding charges, supplemented by the second-order description of $dE/dq$, where $E$ is internal energy and $q$ is electrostatic charge on a given atom. [29,30] The interaction between two charges is written as a shielded Coulomb potential to guarantee correct behavior of covalently bonded atoms. The instantaneous valence force and interaction energy between two atoms are determined by the instantaneous bond order. The latter is determined by the instantaneous bond distance. These interaction energy functions are parametrized vs. QC energy scans involving all applicable types of bond-breaking processes. The bond order concept is used to define other valence interactions, such as bond, lone electron pair, valence angle, conjugation, and torsion angle energies. It is important for energy conservation and stability that all interaction terms smoothly decay to zero during bond dissociation. The conventional pairwise van der Waals energy term describes short-range electron-electron repulsion, preserving atom size, and longer-range London attractive dispersion. Unlike non-reactive MD simulations, ReaxFF uses the van der Waals term for covalently bonded atoms, where it competes with a monotonically attractive bond term. Such an approach to chemical bonding requires a significant number of independent parameters, which can be obtained from QC energies. Bond dissociation, geometry distortion, electrostatic charges, infrared spectra, equations of state and condensed-phase structure are typically derived using an electronic structure method, such as density functional theory, to be consequently used in the ReaxFF parametrization. The works by van Duin, Goddard and coworkers[29,30] provide a more comprehensive description of the methodology used here.

**ACKNOWLEDGMENTS**

MEMPHYS is the Danish National Center of Excellence for Biomembrane Physics. The Center is supported by the Danish National Research Foundation. E. E. F. thanks Brazilian agencies FAPESP and CNPq for support. O.V.P. acknowledges grant CHE-1300118 from the US National Science Foundation.

**AUTHOR INFORMATION**

E-mail addresses for correspondence: vvchaban@gmail.com (V.V.C.); fileti@gmail.com (E.E.F.); prezhdo@rochester.edu (O.V.P.)